\documentstyle[12pt]{article}

\newlength{\extraspace}
\newlength{\extraspaces}
\def\bsklength{.8mm} 

\newcommand{\beq}{\begin{equation}}
\newcommand{\eeq}{\end{equation}}

\newcommand{\beqa}{\begin{eqnarray}}
\newcommand{\eeqa}{\end{eqnarray}}

\newcommand{\newsection}[1]{
\vspace{6mm}
\pagebreak[3]
\addtocounter{section}{1}
\setcounter{equation}{0}
\setcounter{subsection}{0}
\noindent{\large \bf \thesection. #1}
\nopagebreak
\medskip
\nopagebreak}

\newcommand{\newsubsection}[1]{
\vspace{5mm}
\pagebreak[3]

\addtocounter{subsection}{1}
\noindent{ \it \thesubsection. #1}
\nopagebreak
\vspace{2mm}
\nopagebreak}

\def\nonu{\nonumber \\[.5mm]}
\def\half{{\textstyle{1\over 2}}}
\def\pa{\partial}

\def\CD{{\cal D}}

\def\CL{{\cal L}}
\def\CN{{\cal N}}

\def\CO{{\cal O}}

\renewcommand{\hat}{\widehat}
\renewcommand{\tilde}{\widetilde}
\renewcommand{\det}{{\rm det}}

\newcommand{\e}{{\rm e}}

\newfont{\cmss}{cmss10 scaled\magstep1} 
\newfont{\cmsss}{cmss10 }
\def\IZ{\relax\ifmmode\mathchoice
{\hbox{\cmss Z\kern-.4em Z}}{\hbox{\cmss Z\kern-.4em Z}}
{\lower.9pt\hbox{\cmsss Z\kern-.4em Z}}
{\lower1.2pt\hbox{\cmsss Z\kern-.4em Z}}\else{\cmss Z\kern-.4em Z}\fi}
\def\IR{\relax\ifmmode\mathchoice
{\hbox{\cmss I\kern-.5em I}}{\hbox{\cmss R\kern-.5em R}}
{\lower.9pt\hbox{\cmsss I\kern-.5em I}}
{\lower1.2pt\hbox{\cmsss R\kern-.5em R}}\else{\cmss R\kern-.5em R}\fi}
\newcommand{\Tr}{{\rm Tr}}

\def\del{\nabla}
\def\d{{\rm d}}

\begin{document}
\setcounter{page}{0}
\addtolength{\baselineskip}{\bsklength}
\thispagestyle{empty}
\renewcommand{\thefootnote}{\fnsymbol{footnote}}	

\begin{flushright}
{\sc MIT-CTP-2557}\\
hep-th/9608148\\
August 1996
\end{flushright}
\vspace{.4cm}

\begin{center}
{\Large
{\bf{Dilatations Revisited}}}\\[1.2cm] 
{\rm HoSeong La}
\footnote{e-mail address: hsla@mitlns.mit.edu\\			
This work is supported in part by funds provided by the U.S. Department of
Energy (D.O.E.) under cooperative research agreement \#DF-FC02-94ER40818.
 }\\[3mm]							
{\it Center for Theoretical Physics\\[1mm]		
Laboratory for Nuclear Science\\[1mm]
Massachusetts Institute of Technology\\[1mm]
77 Massachusetts Avenue\\[1mm]
Cambridge, MA 02139-4307, USA} \\[1.5cm]

{\parbox{14cm}{
\addtolength{\baselineskip}{\bsklength}
\noindent
Dilatation, i.e. scale, symmetry in the presence of the dilaton in Minkowski
space is derived from diffeomorphism symmetry in curved  spacetime,
incorporating the volume-preserving diffeomorphisms. The conditions for scale
invariance are derived and their relation to conformal invariance is examined.
In the presence of the dilaton scale invariance automatically guarantees
conformal invariance due to diffeomorphism symmetry. Low energy scale-invariant
phenomenological Lagrangians are derived in terms of dilaton-dressed fields,
which are identified as the fields satisfying the usual scaling properties. The
notion of spontaneous scale symmetry breaking is defined in the presence of the
dilaton. In this context, possible phenomenological implications are advocated
and by computing the dilaton mass the idea of PCDC (partially conserved
dilatation current) is further explored.

} 
}\\[1.5cm]


\end{center}
\noindent
\vfill


\setcounter{section}{0}
\setcounter{equation}{0}
\setcounter{footnote}{0}
\renewcommand{\thefootnote}{\arabic{footnote}}	
\newcounter{xxx}
\setlength{\parskip}{2mm}
\addtolength{\baselineskip}{\bsklength}
\newpage

\newsection{Introduction}

Since the MIT-SLAC deep inelastic scattering experiments\cite{dpinel} the role
of dilatations in physics has attracted fair amount of attention. Knowing the
fact that we live in the world of a given scale, the classical scale
symmetry\cite{wess}\cite{rscale}\cite{anom}\cite{spon}  based on the dilatation
of local coordinates in a Lorentz frame is destined to be broken at that given
scale. Scale symmetry breaking in principle can be either explicit or
spontaneous. It turns out that in a simple model with a scalar field the
classical scale symmetry is anomalous at the quantum level, hence it is 
explicitly broken\cite{anom}. In more realistic cases like massless QED or
gauge theories, scale symmetry is also broken by the trace anomaly\cite{racd}.
Nevertheless, there have been attempts to introduce spontaneous  breaking of
scale symmetry with a Goldstone boson called  the {\it dilaton}, as an analog
to the pion for chiral symmetry  breaking\cite{cole}.
In either cases, no significant physical implications  have been understood
except for conformal field theories in  two dimensions\cite{BPZ}.

To further understand the role of scale symmetry in nature, we need to address
the origin of the low energy scale symmetry. It is commonly believed that the
scale symmetry in Minkowski space is an analog to the Weyl symmetry in curved
spacetime. This however is in some sense unsatisfactory because of the lack of
any direct relationship. A low energy scale transformation involves a change of
local coordinates, but a Weyl transformation does not. We need a more direct
connection between properties in curved spacetime and those in a local Lorentz
frame particularly for any spacetime symmetries.  This becomes an important
issue if gravitational effects get stronger. Particularly, in string-motivated
supersymmetric models the dynamics of the dilaton is crucial to understand the
structure of the coupling constants and
supersymmetry\cite{wittdil}\cite{ds}\cite{kaplo}\cite{nilles}.

In this paper we start only with Diff (diffeomorphism) symmetry without
referring to Weyl symmetry, then we shall derive the scale symmetry in
Minkowski space\footnote{In \cite{bd} the relation between the low energy scale
symmetry and the Weyl symmetry in curved spacetime is investigated and the role
of restricted coordinate transformations in this context is suggested.  They
have tried to address the relevance of SDiff symmetry with respect to the
dilaton, but do not succeed  to address subtleties involving SDiff in curved
spacetime, due to using coordinate dependent conditions. The covariant way to
introduce SDiff in curved spacetime is explained in detail by the author in
\cite{sdiffgr}. Nevertheless, there are certain similarities between some
results in \cite{bd} and those of this paper, but careful readers will find the
fundamentals are quite different.}.  Note that in any dimensional spacetime
scale invariance does not necessarily imply Weyl invariance, although it often
implies conformal invariance. This signals that it would be better to
understand scale symmetry as part of  Diff, and that it could provide a natural
explanation of the relation between scale symmetry and conformal invariance.
Diff decomposes into SDiff (volume-preserving diffeomorphisms)  and CDiff
(conformal diffeomorphisms). Since SDiff preserves a volume  element,
dilatations are not part of SDiff. This is the crucial structure to be used in
this paper.

Dilatations are defined in terms of local coordinates by
\beqa
\label{e1}
x &\to & \e^\alpha x ,\\
\label{e2}
\Phi_{[d]}(x) &\to & \e^{d\alpha}\Phi_{[d]}(\e^\alpha x),
\eeqa
where $d$ is the scale dimension (or the conformal weight). Eq.(\ref{e1})
suggests dilatations should be expressed as diffeomorphisms, although
eq.(\ref{e2}) is not a result of a diffeomorphism. We however shall find that
$\Phi_{[d]}$ can be expressed as a dilaton-dressed field and that eq.(\ref{e2})
indeed becomes a result of a diffeomorphism. This can be done only in the 
presence of the dilaton so that in this context scale symmetry naturally 
incorporates the  dilaton. As an important result, scale invariance 
automatically guarantees conformal invariance because both are just part of 
Diff invariance.

Once quantum effects are included, subtraction of ultraviolet divergences 
inevitably  demands the introduction of a renormalization scale as an explicit
mass scale. This input scale is a free parameter and a theory should not depend
on any  changes of such a scale, yet it breaks scale symmetry in a naive sense.
We find that at the quantum level, in the presence of the dilaton, the notion
of the classical scale  symmetry should be generalized to include changes of
the renormalization scale. Then we obtain a conserved generalized scale current
which incorporates naive scale anomalies. There is another anomaly if this
quantum conservation law is not satisfied. As we shall find out, this in fact
is consistent with the idea of the partially conserved dilatation current
(PCDC), whilst the naive scale current is not. This quantum scale current
conservation law produces an analog to the Callan-Symanzik  equation of the
effective potential without explicit variations of coupling constants. 
Then, we can
define the spontaneous breaking of scale symmetry with the dilaton as a 
Goldstone boson, and that in a new symmetry-breaking vacuum,  the PCDC
structure is correctly produced with an anomalous term proportional to square
of the dilaton mass. Without this modification, the leading term of  the usual
anomaly is not related to the dilaton. Therefore, it legitimizes our
generalization.

In four dimensions, spontaneous symmetry breaking can be induced radiatively 
without using anomalies in this sense. It enables us to compute the dilaton 
mass explicitly, hence making the idea of PCDC realistic.  Depending on the
dilaton scale, various physics can be suggested. For example, if the dilaton
scale is  low, the dilaton can be light enough to be a candidate for dark
matter.  Since the dilaton does not couple to gauge fields directly,  but quite
universally couples to fermions and scalars, the existence could be abundant,
yet it could have escaped detections.

There is another implication. The naturalness of mass scales combined with
the dilaton can also explain certain  hierarchies of apparently different mass
scales, with plausible assumptions, because dilaton contribution is often
exponential.

This paper is organized as follows. In section two, dilatations are given in
terms of the Diff in the geometry of 
$g_{\mu\nu} = \e^{2\kappa\phi}\eta_{\mu\nu}$ 
incorporating the SDiff. Compared to Diff, Weyl transformations are not 
accompanied by changes of local coordinates. Then we recover the low energy 
dilaton transformation property. Low energy fields are
in fact dilaton-dressed fields satisfying proper scaling properties, yet they
have correct Lorentz properties. In this context, we can easily obtain the 
dilatation current and the conformal current to check their relationship 
explicitly. We recover the previously known results and the generalization in 
the presence of the dilaton. 
In section three, scale-invariant phenomenological Lagrangians are derived
for various fields including the dilaton and the axion.
In section four, using these phenomenological Lagrangians, we investigate
plausible scenarios of scale symmetry breaking. In particular, introducing
the notion of the spontaneous breaking of scale symmetry based on the 
generalized scale symmetry in the presence of the renormalization scale, we 
could further clarify the idea of PCDC. Also a possibility of explaining certain
(mass) scale hierarchies using the dilaton is presented.
Finally, in the last section, we summarize the results obtained and
some possible future developments are proposed.

\newsection{Defining Dilatations}
\newsubsection{Diff vs. Dilatations}

In curved spacetime, a Weyl transformation is 
\beq
\label{eweyl}
\delta g_{\mu\nu} = 2 \epsilon \rho (x) g_{\mu\nu},
\eeq
where $\rho(x)$ is constant for a global (or rigid) Weyl transformation. 
Usually in literatures this global Weyl transformation is regarded as the
analog to a scale transformation in Minkowski space, hence relating scale
symmetry to Weyl symmetry. But the awkwardness of this relation is that it 
does not naturally lead to the scale symmetry in Minkowski space by simply
taking the flat space limit of curved spacetime. Weyl transformations are 
supposed to be independent from coordinate changes
contrary to scale transformations.

Our motivation is to introduce the scale symmetry in Minkowski space that can
be naturally derived by simply taking the flat space limit of curved spacetime.
Then it enables us to understand the origin of scale symmetry as part of
spacetime symmetries. Furthermore, as soon to be explained, this naturally
introduces the dilaton in Minkowski space and one can investigate 
scale-invariant Lagrangians in this context.

Under Diff, fields transform according to 
\beq
\label{e3}
\left(T_{\mu_1\cdots\mu_p} + \delta T_{\mu_1\cdots\mu_p}\right)
\d x^{\mu_1}\cdots\d x ^{\mu_p} =
T_{\mu_1\cdots\mu_p}(x+\delta x) \d(x+\delta x)^{\mu_1}
\cdots\d(x+\delta x)^{\mu_p}.
\eeq
Then $\delta T_{\mu_1\mu_2\cdots\mu_p}$ is nothing but the Lie derivative
along $\delta x$.
Now consider a metric of the form
\beq
\label{e4}
g_{\mu\nu} = \e^{2\kappa\phi}\eta_{\mu\nu},
\eeq
where $\e^{n\kappa\phi} = \sqrt{g}$ in terms of $g\equiv |\det g_{\mu\nu}|$ and
$\kappa$ is the dilaton scale. The effect of introducing the explicit 
scale parameter, $\kappa$, is to let the dilaton have mass dimension 
$(n-2)/2$, where $n$ is the dimension of spacetime. Note that $\kappa$  is not
really a free parameter because we can always rescale it by rescaling $\phi$.
As far as gravity is concerned, the natural choice of this scale is the Planck
scale. But, here, instead of doing that, we will fix it later at any
phenomenologically proper scale so that we can study the dilaton in an energy
scale much lower than the quantum gravity scale. Since $\kappa$ always appears
in combination with $\phi$, fixing $\kappa$ actually requires a nontrivial
dilaton vacuum expectation value.

As emphasized in ref.\cite{sdiffgr}, if $\phi$ does not transform like a scalar 
under Diff, but the transformation property under Diff is dictated by that of the
metric,  then for $v\equiv \delta x$ in $n$ dimensions 
$\delta g_{\mu\nu} = \nabla_\mu v_\nu + \nabla_\nu v_\mu$ leads to
\beq
\label{e5}
\delta\e^{\kappa\phi} ={\textstyle{1\over n}}\e^{\kappa\phi}D_\mu v^\mu,
\eeq
where 
\beq
\label{eqder}
D_\mu \equiv \pa_\mu + n\kappa\pa_\mu\phi .
\eeq
In terms of $g$, 
$D_\mu \equiv \pa_\mu +\pa_\mu\ln\sqrt{g}$ and
$\pa_\mu\left(\sqrt{g} v^\mu\right) = \sqrt{g} D_\mu v^\mu$.
$D_\mu$ is the same as the covariant derivative $\nabla_\mu$ only
when it acts on a
covariant vector, but, in general, they are different. Eq.(\ref{e5}) shows
that eq.(\ref{e4}) is not to be considered as a conformal gauge fixing 
condition globally, but is a local  expression of a metric in terms of a
non-global function $\phi$. For example, a density is not globally defined 
because it depends on a choice of local coordinates.
If eq.(\ref{e4}) were the 
conformal gauge fixing condition, it would lead to $\delta\e^{2\kappa\phi} =
v^\mu\pa_\mu \e^{2\kappa\phi}$. Under SDiff, $\phi$ behaves like a constant to
make pure $\phi$ Lagrangians manifestly SDiff-invariant.

The consistency condition between eq.(\ref{e5}) and eq.(\ref{e4}) is
\beq
\label{ecnf}
{\textstyle{2\over n}}\eta_{\mu\nu}\pa_\alpha v^\alpha =
\eta_{\mu\alpha}\pa_\nu v^\alpha + \eta_{\alpha\nu}\pa_\mu v^\alpha
\eeq
so that diffeomorphisms of eq.(\ref{e4}) appear as conformal transformations of
flat spacetime.

In particular, for infinitesimal $\alpha$ under the dilatation, eq.(\ref{e1}),
\beq
\label{e6}
\delta\phi = \alpha\left({\textstyle{1\over\kappa}} + x^\mu\pa_\mu\phi\right),
\eeq
which is nothing but the dilatation property given in ref.\cite{cole}.
This shows that the dilatations of the dilaton are results of diffeomorphisms.

Sometimes, it is useful to introduce a field redefinition (without the axion,
see eq.(\ref{e36}) for the definition with the axion,)
\beq
\label{echd1}
\chi \equiv \e^{\kappa\phi}.
\eeq
Under Diff, $\chi$ transforms in a not-so-inspiring way, 
but, under dilatations
\beq
\label{echd2}
\delta\chi = \alpha \left(1 + x^\mu\pa_\mu\right)\chi.
\eeq
Thus, although $\chi$ is not a scalar, it transforms like a scale-dimension-one
field. $\chi$ is mass-dimensionless.

To produce eq.(\ref{e2}) let us introduce a dilaton-dressed field $\Phi_{[d]}$
as
\beq
\label{e7}
\Phi_{[d]} \equiv \e^{d\kappa\phi}\Phi ,
\eeq
where $\Phi$ transforms like a scalar under Diff. This dilaton dressing does
not change the mass dimension of the field. Then under dilatations
\beq
\label{e8}
\delta\Phi_{[d]} = \alpha (d + x^\mu\pa_\mu)\Phi_{[d]}.
\eeq
Such dressing is not needed for vector fields in four dimensions
because under Diff
\beq
\label{e9}
\delta A_\mu = \alpha (1 + x^\lambda\pa_\lambda) A_\mu.
\eeq
Similarly, we can define all dimensional fields in $n$ dimensions by 
properly dressing with the dilaton and the scale transformation properties
follow from the Diff transformation rule. In this sense, the mass dimension of
a field is not necessarily the same as the scale dimension. For example, the
dilaton has mass dimension $(n-2)/2$, but its scale dimension is not even
defined.

One can also easily check that the dilaton is, after all, a Lorentz scalar, 
hence so is $\Phi_{[d]}$. Thus, from the low energy point of view $\Phi_{[d]}$
and $\phi$ are  indistinguishable from the usual scalar field. This clearly
shows that the dilatations in Minkowski space can be derived from the Diff of
virtual spacetime geometry of $g_{\mu\nu} = \e^{2\kappa\phi} \eta_{\mu\nu}$ and
we are never required to introduce Weyl symmetry. 

\newsubsection{Dilatation Current}

The conformal current (of a second order system) can be easily computed as
a Noether current of CDiff such that
\beq
\label{e16}
\hat{J}_{\rm C}^\mu = v^\lambda \tilde{T}^\mu_{\ \lambda} 
+\nabla_\lambda v^\lambda
\tilde{K}^\mu + \nabla_\alpha\nabla_\lambda v^\lambda \hat{L}^{\alpha\mu},
\eeq
where 
$\del_\mu v_\nu +\del_\nu v_\mu = {2\over n}g_{\mu\nu}\del_\alpha v^\alpha$
and $\tilde{T}^\mu_{\ \lambda}$, $\tilde{K}^\mu$, 
$\hat{L}^{\alpha\mu} = \hat{L}^{\mu\alpha}$ are accordingly computed. 
The conformal invariance requires $\del_\mu \hat{J}^\mu_{\rm C} = 0$, which
must  be satisfied if the Lagrangian is Diff invariant. For $g_{\mu\nu} =
\e^{2\kappa\phi}\eta_{\mu\nu}$ and a generic field $\Phi$,
\beqa
\label{e17}
\tilde{T}^\mu_{\ \lambda} &=& \e^{-n\kappa\phi}\left(-\delta^\mu_{\ \lambda}\CL
+{\pa\CL\over \pa(\pa_\mu\Phi)}\pa_\lambda\Phi \right)	\nonu
\tilde{K}^\mu &=& {\textstyle{1\over n\kappa}}\e^{-n\kappa\phi}\left(
{\pa\CL\over \pa(\pa_\mu\phi)}
-\pa_\alpha{\pa\CL\over \pa(\pa_\alpha\pa_\mu\phi)} \right)\\
\hat{L}^{\alpha\mu} &=& {\textstyle{1\over n\kappa}}\e^{-n\kappa\phi}
{\pa\CL\over \pa(\pa_\mu\pa_\alpha\phi)}. 	\nonumber 
\eeqa
In the presence of the dilaton, the conformal current we have obtained 
contains an extra term compared to the one in \cite{wess} and \cite{polch},
which is found by trial and error and also without the dilaton. Here, we have 
derived it as a Noether current with respect to the CDiff so that all
the terms can be computed explicitly once an explicit Lagrangian is given.

To show the extra term explicitly, let us expand the covariant derivative
in terms of the partial derivative to obtain
\beq
\label{e18}
\hat{J}_{\rm C}^\mu =
v^\lambda \hat{T}^\mu_{\ \lambda} +\pa_\lambda v^\lambda\tilde{K}^\mu 
+ \pa_\alpha v^\lambda \left(n\kappa\pa_\lambda \phi \hat{L}^{\alpha\mu}\right)
+ \pa_\alpha\pa_\lambda v^\lambda \hat{L}^{\alpha\mu},
\eeq
where
\beq
\label{e19}
\hat{T}^\mu_{\ \lambda} = \tilde{T}^\mu_{\ \lambda} +n\kappa\pa_\lambda\phi
\tilde{K}^\mu + n\kappa\pa_\alpha\pa_\lambda \phi \hat{L}^{\alpha\mu}.
\eeq
The $\pa_\alpha v^\lambda$ term is extra. As in the flat case, this term can be
rewritten as the second term using the conformal Killing vector condition,
eq.(\ref{ecnf}), only if $\pa_\lambda\phi \hat{L}^{\alpha\mu}$ is symmetric
under exchanging $\alpha$ and $\lambda$ indices.

In our derivation of the conformal current, we assume the presence of 
the second order derivative term of the dilaton, which in fact exists in the
curvature term. If we assume there are no second order derivative
terms of any fields, then $\hat{L}^{\alpha\mu}$ vanishes. However, we shall
find it necessary to keep them to prove conformal invariance.

The dilatation (or scale) current $\hat{S}^\mu$ is $\hat{J}_{\rm C}^\mu$ with 
$v^\lambda = x^\lambda$:
\beq
\label{e20}
\hat{S}^\mu = x^\lambda \hat{T}^\mu_{\ \lambda} + \hat{K}^\mu,
\eeq
where
\beq
\label{e21}
\hat{K}^\mu = n\tilde{K}^\mu + n\kappa\pa_\lambda \phi\hat{L}^{\lambda\mu}.
\eeq
In our case the dilatation current should be covariantly conserved with 
respect to the metric $g_{\mu\nu} = \e^{2\kappa\phi}\eta_{\mu\nu}$. Thus
\beq
\label{e22}
0 = \nabla_\mu \hat{S}^\mu = x^\lambda D_\mu \hat{T}^\mu_{\ \lambda} 
+ \hat{T}^\mu_{\ \mu} + \nabla_\mu \hat{K}^\mu,
\eeq
where $D_\mu$ is introduced in eq.(\ref{eqder}).
This can be satisfied if the stress-energy tensor is conserved as
\beq
\label{eq2}
D_\mu \hat{T}^\mu_{\ \lambda} = 0
\eeq
and 
\beq
\label{eq3}
\hat{T}^\mu_{\ \mu} = - D_\mu \hat{K}^\mu .
\eeq
Note that $\nabla_\mu \hat{T}^\mu_{\ \lambda} \neq 0$. In fact, we should not 
expect that $\nabla_\mu \hat{T}^\mu_{\ \lambda} = 0$ because 
$\hat{T}^\mu_{\ \lambda}$ is derived by choosing a local frame to 
fix coordinates such that $v^\lambda = x^\lambda$. 
On the contrary, the dilatation current $\hat{S}^\mu$ itself is generic so that 
it has to be covariantly conserved and its conservation is dictated by the
Noether's theorem.

Using the property $\sqrt{g} D_\mu \hat{S}^\mu = \pa_\mu S^\mu$, where
$S^\mu \equiv \sqrt{g}\hat{S}^\mu$, the above
conditions can be written as the usual formulas in flat space. All other
tensors relevant in flat space can be similarly defined by multiplying 
$\sqrt{g}$. 

\newsubsection{Conformal Invariance vs. Scale Invariance}

In our case, Diff invariance leads to conformal invariance automatically. More
precisely, conformal invariance requires that the conformal current
$\hat{J}^\mu_{\rm C}$, eq.(\ref{e16}), should be covariantly conserved such
that
\beq
\label{eq4}
0 \!=\! \nabla_\mu \hat{J}^\mu_{\rm C} \!=\! v^\lambda\del_\mu 
\tilde{T}^\mu_{\ \lambda}
+\del_\alpha v^\alpha \left(\half \tilde{T}^\mu_{\ \mu} 
+ \del_\mu\tilde{K}^\mu \right) + \del_\mu\del_\alpha v^\alpha
\left(\tilde{K}^\mu + \del_\lambda \hat{L}^{\mu\lambda}\right) 
+ \del_\mu\del_\lambda\del_\alpha v^\alpha \hat{L}^{\mu\lambda} .
\eeq
In general, the terms in the RHS are not independent so that conformal 
invariance does not necessarily require each term to vanish separately in 
curved spacetime. In a generic second order system, $\hat{L}^{\mu\lambda} =
g^{\mu\lambda} \hat{L}/n$. For this reason,  one might be tempted to demand
$\del_\alpha v^\alpha$ is harmonic and each term vanishes separately. But, this
does not happen. One obvious reason is that  $\tilde{T}^\mu_{\ \lambda}$ is not
necessarily a covariantly conserved stress-energy tensor in curved spacetime.

For our purpose, as long as the conformal current is covariantly conserved,
we do not need covariant conditions on other terms because we are interested
in conditions to be imposed in a local Lorentz frame. Nevertheless, since 
any metric can be written as $g_{\mu\nu} = \e^{2\kappa\phi}\eta_{\mu\nu}$ for
nonscalar $\phi$, we can still retain all necessary geometric 
data\cite{sdiffgr}.

In a local Lorentz frame we can choose coordinates such that $v^\lambda$,
$\pa_\alpha v^\lambda$, $\pa_\alpha\pa_\beta v^\lambda$...etc., are linearly
independent. Then
\beqa
\label{eq5}
0 = \nabla_\mu \hat{J}^\mu_{\rm C} &\!=\!& v^\lambda D_\mu 
\hat{T}^\mu_{\ \lambda} \nonu
&&\!\!\!\!+\pa_\mu v^\lambda \left(\hat{T}^\mu_{\ \lambda} 
+ D_\alpha\left(\delta^\mu_{\ \lambda}\tilde{K}^\alpha 
+\pa_\lambda\ln\sqrt{g}\, \hat{L}^{\alpha\mu}\right)\right) \nonu
&&\!\!\!\!+ \pa_\mu\pa_\sigma v^\lambda \left(\delta^\sigma_{\ \lambda}
\left(\tilde{K}^\mu + D_\alpha \hat{L}^{\alpha\mu}\right) 
+\pa_\lambda\ln\sqrt{g}\, \hat{L}^{\sigma\mu}\right) \nonu
&&\!\!\!\!+ \pa_\mu\pa_\lambda\pa_\alpha v^\alpha \hat{L}^{\sigma\mu} 
\eeqa
implies the four terms in the RHS vanish independently, leading to
\beqa
\label{eq6}
 &&0 =D_\mu \hat{T}^\mu_{\ \lambda},	\\
\label{eq7}
&&0=\pa_\mu v^\lambda \left(\hat{T}^\mu_{\ \lambda} 
+ D_\alpha\left(\delta^\mu_{\ \lambda}\tilde{K}^\alpha 
+\pa_\lambda\ln\sqrt{g}\, \hat{L}^{\alpha\mu}\right)\right),	\\
\label{eq8}
&&0=\pa_\mu\pa_\sigma v^\lambda \left(\delta^\sigma_{\ \lambda}
\left(\tilde{K}^\mu + D_\alpha \hat{L}^{\alpha\mu}\right) 
+\pa_\lambda\ln\sqrt{g}\, \hat{L}^{\sigma\mu}\right), \\
\label{eq9}
&&0=\pa_\mu\pa_\lambda\pa_\alpha v^\alpha \hat{L}^{\sigma\mu} .
\eeqa
Eq.(\ref{ecnf}) implies that eq.(\ref{eq9}) is true if $\hat{L}^{\mu\lambda} =
g^{\mu\lambda} \hat{L}/n$ for $g_{\mu\nu} = \e^{2\kappa\phi}\eta_{\mu\nu}$.
Note that in eqs.(\ref{eq7})(\ref{eq8}) we require the whole thing to vanish
instead of the formulae inside the bracket. It is because these conditions in 
fact depend on the specific details of $v^\lambda$. This is one of the reasons
the conformal structure actually depends on the dimensionality of spacetime.

In Euclidean two dimensions, the second term in eq.(\ref{eq8}) also vanishes
because $v^\lambda$ itself is harmonic. Then, using the trace condition
eq.(\ref{eq3}), one can  show that eq.(\ref{eq7}) is satisfied. The only
remaining condition is 
\beq
\label{eq10}
0 = \tilde{K}^\mu + D_\alpha \hat{L}^{\alpha\mu}.
\eeq
In Minkowski two dimensions, although we cannot use the power of complex
analysis, yet the Diff invariance dictates $v^\lambda$ should be still
harmonic. In fact, as far as physics is concerned, we can impose the
light-front conditions to derive equivalent conditions. Therefore, in two
dimensions for scale symmetry to imply conformal invariance, one extra
condition eq.(\ref{eq10}) needs to be satisfied. Expressing in the flat space
objects, this is the same condition as the one given by Polchinski\cite{polch}. 

In other $n$ dimensions, one can use explicit special conformal transformations
\beq
\label{eq11}
v^\lambda = a_\alpha ( - \eta^{\alpha\lambda} x^2 + 2 x^\alpha x^\lambda).
\eeq
Regardlessly of any $\hat{L}^{\sigma\mu}$, eq.(\ref{eq9}) is true and
eq.(\ref{eq7}) generally follows from scale invariance.
The extra condition, eq.(\ref{eq8}), now reads
\beq
\label{eq12}
n\left(\tilde{K}^\mu + D_\alpha \hat{L}^{\mu\alpha}\right) + 2
\pa_\alpha\ln\sqrt{g}\hat{L}^{\mu\alpha} -
\eta^{\mu\nu}\pa_\nu\ln\sqrt{g}\eta_{\lambda\alpha}\hat{L}^{\lambda\alpha} = 0.
\eeq
Due to the dilaton the extra terms compared to eq.(\ref{eq10}) do not vanish.
For $\hat{L}^{\mu\lambda} = g^{\mu\lambda} \hat{L}/n$ the extra terms contain 
a prefactor $(2-n)$ so that they vanish only in two dimensions.
Note that, as long as the action is Diff invariant, eq.(\ref{eq10})
and eq.(\ref{eq12}) are supposed to be satisfied automatically. This can be
achieved because of the dilaton. 

This in turn implies that scale symmetry breaking amounts Diff symmetry
breaking of $g_{\mu\nu} = \e^{2\kappa\phi}\eta_{\mu\nu}$ and remaining symmetry
is supposed to be SDiff symmetry, although only Poincar\'e symmetry is usually
manifest in flat spacetime due to a gauge choice.

\newsection{Scale-invariant Phenomenological Lagrangians}

Scale-invariant Lagrangians are nothing but the usual Lagrangians of tensor
fields defined in the curved spacetime of the dilaton virtual geometry. Then we
simply express them in terms of the dilaton-dressed fields.

\newsubsection{Dilaton}

The kinetic energy for the dilaton in Minkowski space
can be derived from the curvature of the
virtual geometry of metric $g_{\mu\nu} = \e^{2\kappa\phi}\eta_{\mu\nu}$. 
Let us first consider the stringy Lagrangian
$$\CL_{{\rm S}} = -{\textstyle{1\over 2\kappa^2}}
\sqrt{g}\e^{-2\Phi}(R - 4 g^{\mu\nu}\pa_\mu\Phi\pa_\nu\Phi).$$
One can quickly check that $\phi \propto \Phi$ does not lead to a scale 
invariant action for the above metric.  Although it is possible\cite{cnfdil},
here we do not want to have another  gravitational scalar field, so we shall
seek a case without one. On the contrary, the only scale invariant Lagrangian
we can obtain is  to let $g_{\mu\nu} = \eta_{\mu\nu}$ and then 
$\Phi = -{n\over 2}\phi$. 
In other words, in the string frame once the dilaton is identified
separately\cite{ftdil},  the metric should not contain any dilaton degrees of
freedom so that the metric should be taken to be flat for our purpose.  Thus
the stringy dilaton and the low energy dilaton become equivalent only if we
incorporate SDiff. Anyhow, this is equivalent to choosing the Einstein-Hilbert
Lagrangian 
\beq
\label{e10}
\CL_{{\rm EH}} = -\sqrt{g} R/\CN\kappa^2
\eeq 
where $\CN$ is a normalization numerical prefactor. At the scale in which the
gravity is relevant, one should choose $\CN = 2$ and $\kappa^2 = 8\pi G$. But,
here we shall choose them to make the kinetic energy term canonical.
It also confirms that in our case scale symmetry is not related to the global 
Weyl symmetry because $\CL_{{\rm EH}}$ is not Weyl invariant if $n\neq 2$.

For the given metric in $n$ dimensions ($n\neq 2$)
\beq
\label{e11}
\CL_{{\rm EH}} = -{\textstyle{1\over \CN}}
(n-1)(n-2)\e^{(n-2)\kappa\phi}\eta^{\mu\nu}\pa_\mu\phi\pa_\nu\phi
+ {\rm total\ \ derivative}
\eeq
In case of proving conformal invariance the total derivative term is important.
Note that it has an analogous form to the nonlinear chiral Lagrangian except 
the prefactor of $\phi$ is real.

In two dimensions $\CL_{{\rm EH}}$ does not provide the dilaton kinetic 
energy. Instead, we could use
\beq
\label{elv}
\CL_{{\rm L}} = -{\textstyle{1\over 8}}\sqrt{g} R \Delta^{-1} R
= \half\kappa^2\eta^{\mu\nu}\phi\pa_\mu\pa_\nu\phi,
\eeq
which is nothing but the Liouville Lagrangian. In two dimensions it is more 
natural to absorb $\kappa$ into $\phi$ so that $\phi$ becomes
mass-dimensionless.

The main difficulty of understanding the dilaton dynamics lies on the
undesirable structure of its potential energy.  From the curved spacetime point
of view the only allowed tree level potential is the exponential one, hence
there is no stable dilaton vacuum. The well known difficulty of handling such a
case is typified in the Liouville theory in two
dimensions\cite{liou}\cite{glg}\cite{sdiffgr}. This is also partly a source of
the ``runaway dilaton problem'' in supergravity models\cite{ds}. Without such a
tree level potential, the dilaton generates a so-called flat direction along
which vacua are degenerate.

As alluded in refs.\cite{sdiffgr}\cite{cnfdil},  we expect that this difficulty
might be overcome if we abandon the manifest Diff invariance and rely on SDiff
invariance only. Without Diff, at least the symmetry alone does not forbid
other stabilizing potentials. In fact careful observation reveals that the
dilaton $\phi$ is not even an ordinary scalar field as shown in eq.(\ref{e5}).
In our case, being in a local Lorentz frame, we can accommodate this structure
naturally for the low energy dilaton. Then, in the next section,  we shall
derive the effective potential that indeed has a symmetry breaking vacuum.

\newsubsection{Scalar Fields}

The scale invariant lagrangian for a scalar field is
\beq
\label{e13}
\CL_\Phi = -\half \eta^{\mu\nu}
\left(\pa_\mu - d\kappa\pa_\mu\phi\right)\Phi_{[d]}
\left(\pa_\nu - d\kappa\pa_\nu\phi\right)\Phi_{[d]}
	-\half m^2 \e^{2\kappa\phi} \Phi_{[d]}^2 -\cdots,
\eeq
where $\Phi_{[d]}$ the dilaton-dressed scalar
field (eq.(\ref{e7})) and $d\equiv (n - 2)/2$, which is nothing but the
mass dimension of the dilaton\footnote{This scale invariant Lagrangian is also
derived in \cite{bd} and the relevance of the conformally flat metric,
eq.(\ref{e4}), is noticed.}. 
$\kappa\pa_\mu\phi$ acts like minimal coupling of a gauge field with a charge
$id$ and the dilaton-scalar couplings are derivative. 
In two dimensions, in particular, this dilaton coupling disappears so that
the scalar field does not interact with the dilaton. In other dimensions the
dilaton-scalar couplings are nonrenormalizable, hence the Lagrangian appears as
an effective field theory Lagrangian\cite{reft}.

\newsubsection{Fermions}

Similarly, in terms of dilaton-dressed fermionic fields $\psi_{[d']} \equiv
\e^{d'\kappa\phi}\psi$ with $d' \equiv (n-1)/2$,
\beq
\label{e14}
\CL_\psi = i\overline{\psi}_{[d']}\gamma^\mu\left(\pa_\mu - d'\kappa\pa_\mu\phi
\right)\psi_{[d']} - m' \e^{\kappa\phi}\overline{\psi}_{[d']}\psi_{[d']},
\eeq
where the gamma matrices are those in Minkowski space. 
The dilaton-fermion coupling is also derivative and non-renormalizable.
The dilaton interacts with fermions in any $d>1$ dimensions.

\newsubsection{Gauge Fields}

We shall always choose gauge fields to be scale-dimension one so that they 
are not dressed by the dilaton. In this way, we can relate the dilaton with 
the gauge coupling constant.
\beq
\label{e15}
\CL_{{\rm YM}} = - {\textstyle{1\over 2g^2}}\e^{(n-4)\kappa\phi}
\eta^{\mu\lambda}\eta^{\nu\sigma} \Tr F_{\mu\lambda} F_{\nu\sigma}.
\eeq
In four dimensions, the dilaton and gauge fields decouple.
In other than four dimensions, the effective gauge coupling constant depends on
the dilaton as in string motivated models.

\newsubsection{Axion}

As a matter of fact, there is another way to define $\chi$ than 
eq.(\ref{echd1}). Simply allow $\chi$ to be complex such that 
$\chi^*\chi = \e^{2\kappa\phi}$. Then, there is an 
undetermined phase $a$, which we shall call the axion, so that
\beq
\label{e36}
\chi \equiv \e^{\kappa(\phi + ia)}.
\eeq
Under Diff, this axion transforms like a (pseudo-)scalar so that, in
particular, under dilatations, $\chi$ still transforms according to
eq.(\ref{echd2}).

A real scalar field should be still dressed as eq.(\ref{e7}), but a charged 
complex field now can be dressed by the complex $\chi$. This in fact is
consistent with the low energy axion associated with U$(1)_{{\rm PQ}}$ because
the axion is only associated with a charged scalar. Despite this fact,
note that the axion we defined here is not the low energy axion\cite{raxion}.
Being part of the geometrical data,
it is rather the gravitational axion associated with the dilaton\cite{wittdil}.
Nevertheless, upon dressing, this axion mixes with the low energy axion
so that it might not be distinguishable in practice.

The metric being real, the kinetic term of the axion cannot be derived from
the curvature term. Thus the axion kinetic term can be introduced by hand as
\beq
\label{e37}
\CL_a = -{\textstyle{1\over \CN}}
 \e^{(n-2)\kappa\phi}\eta^{\mu\nu}\pa_\mu a\pa_\nu a,
\eeq
where $\CN ={2\over (n-2)^2}$ if $n\neq 2$ and $\CN = 2$ if $n=2$.
Combined with $\CL_\phi$ that can be read off from $\CL_{{\rm EH}}$ or
$\CL_{\rm L}$, we can obtain
\beq
\label{e38}
\CL_\chi \equiv \CL_\phi + \CL_a = 
-{\textstyle{1\over 2\kappa^2}}
\eta^{\mu\nu}\pa_\mu\chi_d^*\pa_\nu\chi_d,
\eeq
where $\chi_d \equiv \chi^d$ for $d = (n-2)/2$ if $n\neq 2$ and 
$\chi_d \equiv \kappa(\phi + ia)$ for $d=1$ if $n=2$.
In four dimensions the axion term can be naturally derived from antisymmetric
field $H_{\mu\nu\lambda}$ such that $\pa_\sigma a = 
(1/6)\epsilon_{\sigma\mu\nu\lambda} H^{\mu\nu\lambda}$\cite{wittdil}.

In this case, the dilaton-dressed charged scalar field is
\beq
\label{e39}
\Phi_{[d]} = \chi\,\Phi
\eeq  
so that the scale-invariant scalar Lagrangian coupled to gauge fields becomes
\beq
\label{e40}
\CL_\Phi = -\eta^{\mu\nu}\left(\CD_\mu\Phi_{[d]}\right)^\dagger
\CD_\nu\Phi_{[d]} ,
\eeq
where $\CD_\mu \equiv\left[\pa_\mu -igA_\mu-d\kappa\pa_\mu(\phi +ia)\right]$.
Unlike the dilaton, in the presence of U(1) gauge field the axion can be
gauged away if the global part of U(1) is of Peccei-Quinn type. 
This means that the presence of the axion breaks not only U(1)$_{{\rm PQ}}$
but also this U(1). To avoid breaking the local U(1) one can always insist
the axion does not depend on the U(1) gauge degrees of freedom, but accidental
alignment of the axion direction and U(1) gauge degrees may not be unavoidable.
For other gauge fields axion couplings are
given according to properly modified $\CD_\mu$. 
For example, for fermions in eq.(\ref{e14})  
$(\pa_\mu -d'\kappa\pa_\mu\phi)$ should be replaced with $\CD_\mu$ which axial
axion coupling, which can be done by dressing Weyl fermions with the complex 
$\chi$.

\newsubsection{Example: Dilaton-Scalar System}

Let us demonstrate the previous symmetries for the dilaton-scalar Lagrangian in
four dimensions:
\beq
\label{e23}
\CL = {1\over 2\kappa}\e^{2\kappa\phi}\eta^{\mu\nu}\pa_\mu\pa_\nu\phi
+\half \e^{2\kappa\phi}\eta^{\mu\nu}\pa_\mu\phi\pa_\nu\phi
-\half\eta^{\mu\nu}
\left(\pa_\mu -\kappa\pa_\mu\phi \right)\Phi_{[1]}
\left(\pa_\nu -\kappa\pa_\nu\phi \right)\Phi_{[1]}.
\eeq
We explicitly restored the second order derivative term for conformal
invariance. Actual computation is easier in terms of $\Phi =
\e^{-\kappa\phi}\Phi_{[1]}$, but the result is equivalent. The equations of
motion are 
\beqa
\label{e24}
&&0= \eta^{\mu\nu} (\pa_\mu +2\kappa\pa_\mu\phi)\pa_\nu\Phi \\
\label{e25}
&&0=\eta^{\mu\nu}\left(\pa_\mu\Phi\pa_\nu\Phi 
-{\textstyle{1\over\kappa}}(\pa_\mu + \kappa\pa_\mu\phi)\pa_\nu\phi\right).
\eeqa
Relevant operators are
\beqa
\label{e26}
\tilde{K}^\mu \!&=&\! 0,\quad
\hat{L}^{\mu\nu} = {\textstyle{1\over 8\kappa^2}}
\e^{-2\kappa\phi}\eta^{\mu\nu} 
\nonu
\hat{T}^{\mu}_{\ \lambda} \!&=&\! \e^{-2\kappa\phi}\eta^{\mu\nu}\left(
{\textstyle{1\over 2\kappa}}\pa_\nu\pa_\lambda\phi
-\pa_\nu\Phi\pa_\lambda\Phi\right).
\eeqa
Then
\beqa
\label{e27}
&&D_\mu \hat{T}^\mu_{\ \lambda} 
= (\pa_\mu +4\kappa\pa_\mu\phi)\hat{T}^\mu_{\ \lambda} =0,\\
\label{e28}
&&\hat{T}^\mu_{\ \mu} + D_\mu \hat{K}^\mu = 0
\eeqa
are easily satisfied so that $\CL$ is scale invariant. Using eq.(\ref{eq12}),
it can be shown that for the given $\hat{L}^{\mu\nu}$
the extra condition for $\CL$ to be conformally invariant
becomes simply $\tilde{K}^\mu = 0$, which is indeed satisfied in
eq.(\ref{e26}). Without the second order derivative term in eq.(\ref{e23}),
$\CL$ does not lead to a conserved conformal current. Despite that two
Lagrangians determine the same scale-invariant classical theory if the
difference is a divergence, only one leads to the correct conserved conformal
current. This is because the divergence term is necessary for the dilaton to be
Diff-invariant. It also indicates that equations of motion are more fundamental
than a Lagrangian, as far as symmetries are concerned\footnote{For some 
description of symmetry structure of equations of motion in general and
applications, see ref.\cite{mysym}.}. 
Total divergence terms in a Lagrangian will be automatically taken care of 
in the symmetry analysis of equations of motion.

\newsection{Scale Symmetry Breaking}

Scale symmetry is a continuous global spacetime symmetry based on rigid 
transformations. It is natural to expect that the symmetry will be broken to 
provide us a scale to live on.  As we can see below, there are many different
ways to break scale symmetry. The most common case is that scale symmetry is
explicitly broken because it is anomalous due to quantum effects. This happens
because of the appearance of the renormalization scale. Exceptional cases  can
occur only if beta-functions of coupling constants vanish. But in the presence
of the dilaton, this is the least interesting case simply because it does not
allow any particles associated with the symmetry breaking. Thus we shall not
consider explicit breaking cases here. Having the dilaton means that scale
symmetry is presumed to be spontaneously broken.  From here on, we shall focus
on four dimensional cases, unless specified otherwise.

\newsubsection{Spontaneous Scale Symmetry Breaking}

The scale invariant Lagrangians we derived in the previous section already
define effective field theories  because nonrenormalizable higher order dilaton
terms are involved. The potential term is not yet introduced. One can
explicitly introduce  scale-violating dilaton potential to provide the dilaton
mass\cite{cole}, but here we would like to do without explicitly introducing
symmetry breaking  dilaton potentials. As we shall soon find out, the 1PI
effective potential,  generated from the typical scale-invariant tree level
dilaton potential, induces spontaneous breaking of scale symmetry\footnote{The
subtleties involving the derivation of 1PI effective potential from an
(Wilsonian) effective field theory, which is nonrenormalizable, will be
addressed later.}. Despite that the symmetry in consideration is a continuous
one, the dilaton becomes massive in the new vacuum. This is not really unusual 
because, unlike in other cases with continuous internal
symmetries,  there is only one
real field is involved. So it imitates the case of spontaneous breaking of
a discrete symmetry, which does not involve massless Goldstone bosons. Perhaps,
this indicates the dilaton is in fact part of the graviton and scale symmetry
breaking as part of Diff contains an analog to the Higgs mechanism to make the
dilaton massive as advocated in  ref.\cite{cnfdil}. Thus the 1PI effective
potential simply shows the result of this Higgs mechanism and the dilaton has
eaten up an implicit Goldstone boson.

Before we begin the actual computation, let us first clarify what we mean by
``spontaneous'' scale symmetry breaking here, since we intend to derive this 
breaking radiatively. Any renormalization of an ultraviolet
divergent process introduces a
subtraction (or, renormalization) scale as an explicit mass scale, hence
breaking scale symmetry in a naive sense. Unless the beta function vanishes, 
the effect of such breaking appears as an anomaly. As we mentioned before, if
an anomaly breaks scale symmetry, the introduction of the dilaton is
meaningless. Thus, in the presence of the dilaton, we need to carefully 
understand the meaning of the subtraction scale.

As we shall see, both 1PI effective potentials for the dilaton and a scalar,
$V_{\phi,{\rm eff}}$ and $V_{\Phi,{\rm eff}}$ respectively, are scale
invariant, if we allow the subtraction scale changes under infinitesimal scale 
transformations eq.(\ref{e1}) as
\beq
\label{e30}
\delta M = \alpha M.
\eeq
While $V_{\Phi,{\rm eff}}$ still has a scale
invariant invariant vacuum,
$V_{\phi,{\rm eff}}$ no longer has a scale invariant vacuum. Thus, including
eq.(\ref{e30}) as part of the scale transformation makes the effective action
preserve scale symmetry in the presence of the dilaton as it should be, yet
there is a new vacuum in which this ``generalized'' scale symmetry is
spontaneously broken. Furthermore, this is also perfectly consistent with the
idea of PCDC as we shall show later.

Consider the pure dilaton Lagrangian in four dimensions given by
\beq
\label{e31}
\CL_\phi = -\half\e^{2\kappa\phi} 
\eta^{\mu\nu}\pa_\mu\phi\pa_\nu\phi - {\Lambda\over 4}\e^{4\kappa\phi}.
\eeq
This is a four dimensional analog to the two dimensional 
Liouville Lagrangian in the sense that eq.(\ref{e11}) with the cosmological 
constant term for $n=2$ is the Liouville Lagrangian if $\CN = 2(n-1)(n-2)$.
(Or, more precisely, eq.(\ref{elv}).)
The presence of the prefactor $\e^{2\kappa\phi}$ can be absorbed into the 
derivative by $\chi \equiv \e^{\kappa\phi}$ so that we have some resemblance
to the nonlinear chiral Lagrangian of pions except $\chi$ here is real and
contains only one real field. 

To compute the effective potential, first of all, we need to locate a vacuum.
Since the derivative terms do not contribute in determining a vacuum,
$\CL_\phi$ does not have a stable vacuum, but only a runaway vacuum. In
practice, we can still use the asymptotic vacuum, which is the limit of the
runaway vacuum and technically stable, to compute the effective potential. This
is good enough because the functional integral in fact
only requires an asymptotic
vacuum. Also, for an effective field theory, nonrenormalizable terms are
irrelevant particularly in the low momentum regime, we can still compute the
effective potential for renormalizable interactions.  Even in the high momentum
regime, if the effective field theory for the dilaton is derived from a
well-defined renormalizable theory in the quantum gravity scale, we can still
have plenty of counter terms to tame all the higher order  terms in the
functional integral. In this sense, the square term that is needed for the 1PI
effective potential can be assumed to be fairly unambiguous.

Note that eq.(\ref{e31}) indeed has an asymptotic vacuum that is the limit
of the long tail, since the potential is bounded below. So we compute
the 1PI dilaton effective potential to obtain
\beq
\label{e32}
V_{\phi,{\rm eff}} = {\Lambda\over 4}\e^{4\kappa\phi} +
{1\over 64\pi^2}\left(4\kappa^2\Lambda\e^{2\kappa\phi}\right)^2
\left(\log{4\kappa^2\Lambda\e^{2\kappa\phi}\over M^2} 
-{\textstyle{3\over 2}}\right).
\eeq
One can easily check that this effective potential is scale invariant
incorporating eq.(\ref{e30}). Without eq.(\ref{e30}), the 
$\phi\e^{4\kappa\phi}$ term is not scale invariant.

$V_{\phi,{\rm eff}}$ has a new minimum located at
\beq
\label{e33}
\langle\phi\rangle = {1\over 2\kappa}\left(1 - {\pi^2\over \kappa^4\Lambda}
+\log{M^2\over 4\kappa^2\Lambda}\right)
\eeq
for any $\kappa^4\Lambda$.
And in this new vacuum the scale symmetry is spontaneously broken in the sense
we explained before. In fact the derivative term in eq.(\ref{e31}) 
also breaks the generalized scale symmetry after shifting to the new vacuum. 
In the context of 
$g_{\mu\nu} = \e^{2\kappa\phi}\eta_{\mu\nu}$, this corresponds to spontaneous 
symmetry breaking of Diff to SDiff, as advocated in ref.\cite{cnfdil}.

This spontaneous breaking of scale symmetry can also be elegantly described in
terms of an analog to the Callan-Symanzik equation of the effective potential.
Without introducing the renormalized coupling constant, under the generalized
scale  invariance the invariant effective potential satisfies 
\beq
\label{ec1}
\left(M{\pa\over\pa M} +{1\over \kappa}{\pa\over\pa\phi} 
+d\Phi_{[d]} {\pa\over\pa\Phi_{[d]}} -n\right) 
V_{{\rm eff}}(M,\phi,\Phi_{[d]}) = 0.
\eeq
Note that an anomaly with respect to the usual scale symmetry 
corresponds to 
\beq
\label{ec2}
\pa_\mu S^\mu = -M{\pa V_{{\rm eff}}\over\pa M}.
\eeq
But this is absorbed into the equation and we would not count it as an anomaly
under the generalized scale symmetry. After shifting the vacuum, when the
equation is rewritten in terms of new vairables, if there is a
term violating this form of an equation, 
the scale symmetry is spontaneously broken. As
one can easily check, if the dilaton effective potential has a new vacuum with
a nontrivial vacuum expectation value, then the above equation is not satisfied
after shifting the vacuum because of additional term
$M(\pa\langle\phi\rangle/\pa M)(\pa V_{{\rm eff}}/\pa\phi)$. This will appear
as anomalous dilatation current conservation law
\beq
\label{ec3}
\pa_\mu S_{{\rm Q}}(M,\phi) 
= -M{\pa\langle\phi\rangle\over\pa M}
{\pa V_{{\rm eff}}\over\pa\phi}\Bigg|_{\phi +\langle\phi\rangle},
\eeq
where $S_{{\rm Q}}$ is the generalized dilatation current.

\newsubsection{Via Internal Symmetry Breaking: Without Dilaton Loops}

We can also break scale symmetry in connection with the spontaneous symmetry
breaking of internal symmetries. This can be done because the latter
involves a vacuum expectation value of a field,
which happens to fix a scale. Thus scale symmetry should be expected to be
broken at the same time.

For the argument's sake, let us consider the dilaton-scalar system with $\IZ_2$ 
symmetry and introduce internal-symmetry breaking term
\beq
\label{e34}
V(\Phi_{[1]}) = {\lambda\over 4} \left(\Phi_{[1]}^2 - v^2\right)^2.
\eeq
This tree level scalar potential contains explicit scale symmetry breaking
terms: $\lambda v^4$ and $\lambda v^2\Phi_{[1]}^2$. Thus, scale symmetry
is explicitly broken by hand.
Perturbed around a new vacuum, the potential yields terms that break both 
internal symmetry and scale symmetry.

We might ask what happens to the dilaton, since scale symmetry is broken. Note
that the kinetic energy term of the scalar field leads to a term which modifies
the dilaton kinetic energy term  so that the dilaton Lagrangian, eq(\ref{e31}),
now reads
\beq
\label{e35}
\tilde{\CL}_\phi = -\half\left(\e^{2\kappa\phi} +v^2\kappa^2\right)
\eta^{\mu\nu}\pa_\mu\phi\pa_\nu\phi - {\Lambda\over 4}\e^{4\kappa\phi}.
\eeq
By shifting $\phi$, $v^2\kappa^2$ term cannot be removed. Thus,
the vacuum is still a runaway vacuum, despite the fact that scale symmetry
is broken. This is not unexpected. Since scale symmetry is explicitly broken,
there is no reason to introduce the dilaton from the beginning. As a matter of
fact, near the runaway vacuum the dilaton decouples from the scalar system.
The situation does not become better for field redefinition 
$\chi \equiv \e^{\kappa\phi}$ to allow $\chi$ to shift instead of $\phi$. 

For a continuous internal symmetry, say, U(1), we can introduce the axion
according to the previous section. Due to the axion-scalar coupling, upon
breaking of U(1), the axion kinetic energy can be modified too. For the
dilaton-axion kinetic energy, eq.(\ref{e38}),  any modification of the
prefactor of the dilaton kinetic energy must be accompanied by that of the
axion.  And the prefactor cannot be removed by shifting $\chi$. Thus, it is not
necessary for the dilaton to obtain a nontrivial vacuum  expectation value if
scale symmetry breaking is due to an explicit scale symmetry breaking term in a
Lagrangian.

Let us next check a case without putting an explicit symmetry breaking term.
For a tree level dilaton-scalar coupling without derivatives, we can introduce 
a simple scale invariant term 
\beq
\label{eq15}
-\half m^2\e^{2\kappa\phi}\Phi^2_{[1]}.
\eeq
Combined with the tree level dilaton potential, the potential in this case
reads
\beq
\label{e43}
V(\phi,\Phi) = {\lambda\over 4}
\left(\Phi_{[1]}^2 - v^2\e^{2\kappa\phi}\right)^2 
+{\textstyle{1\over 4}}\left(\Lambda -\lambda v^4\right)\e^{4\kappa\phi}.
\eeq
The vacuum conditions are $\Phi_{[1]}^2 = v^2\e^{2\kappa\phi}$ and
$\e^{2\kappa\phi} = 0$, hence there is no spontaneous breaking of internal 
symmetry. This symmetry breaking can occur only if the dilaton gets  a
nontrivial vacuum expectation value, that is, scale symmetry is also broken.
Therefore, we need to introduce explicit scale symmetry breaking term for tree
level symmetry breaking. Furthermore, to break scale symmetry at the same
time to obtain a nontrivial dilaton vacuum expectation value we need polynomial
terms of $\phi$. 

The radiative case is slightly different. Let us consider scale-invariant
scalar self-interaction $\lambda\Phi^4_{[1]}$ without a mass term. In four
dimensions, no explicit dilaton is involved in this interaction. Then one-loop
contributions of the scalar field generate a double well effective potential
with a new vacuum\cite{colwein}. Perturbed around this new vacuum, 
$\Phi_{[1]} = v + \varphi$, the effective potential now contains a term  
$\lambda v^2\varphi^2$. The presence of such a term breaks the classical scale
symmetry, which is the case with $\delta M = 0$. In fact, the classical scale
symmetry is broken by any term other than $\Phi_{[1]}^4$.
Under the generalized scale symmetry, one can easily check that the effective
potential is invariant both before and after shifting the $\Phi_{[1]}$ vacuum.

\newsubsection{Simultaneous Symmetry Breaking: The Internal and Scale}

In the previous section, 
we noticed that internal symmetry breaking in the scalar sector 
alone does not generate a nontrivial dilaton potential to fix the dilaton
vacuum expectation value, despite that  apparent scale symmetry breaking
occurs. This inconsistency simply tells us that such a semiclassical way of
deriving internal symmetry breaking in the presence of the dilaton is not
correct. The dilaton should be accounted in the same way.

For this purpose, we compute the full effective action
\beq
\label{eq16}
V_{{\rm eff}}(\phi,\Phi) = V_{\phi,{\rm eff}} + V_{\Phi,{\rm eff}},
\eeq
where, in the one scalar case, $V_{\Phi,{\rm eff}}$ is the usual 
Coleman-Weinberg effective potential
\beq
\label{eq17}
V_{\Phi,{\rm eff}} = {\lambda\over 4}\Phi^4_{[1]} + {1\over 64\pi^2}
\left(3\lambda\Phi^2_{[1]}\right)^2
\left(\log{3\lambda\Phi^2_{[1]}\over M^2}-{\textstyle{3\over 2}}\right)
\eeq
and $V_{\phi,{\rm eff}}$ is given in eq.(\ref{e32}). There is no $\phi$-$\Phi$
mixed term in the effective potential because dilaton-scalar couplings in the
tree Lagrangian are all derivative ones.

In this case, the vacuum expectation values of the scalar and the dilaton are
related in terms of the common renormalization scale $M$ so that
\beq
\label{eq18}
\langle\phi\rangle = {1\over 2\kappa}\left({16\pi^2\over 9\lambda} 
-{\pi^2\over \kappa^4\Lambda} + 
\log{3\lambda\langle\Phi_{[1]}\rangle^2\over 4\kappa^2\Lambda}\right).
\eeq
Thus, scale symmetry as well as internal symmetry are broken at the same time.

The naturalness, which means that all the dimensionful parameters in a theory
should be of the same order, implies $\Lambda^{1/4} \sim
\langle\Phi_{[1]}\rangle \sim\langle\phi\rangle$. This in turn implies 
\beq
\label{eq19}
\kappa^4\Lambda \sim \kappa^2\langle\Phi_{[1]}\rangle^2 
\sim \kappa\langle\phi\rangle.
\eeq
For eq.(\ref{eq18}), the above is consistent with
\beq
\label{eq20}
\langle\phi\rangle \sim \CO(1/\kappa).
\eeq
Thus in the new vacuum the dilaton scale $\kappa$ determines the scale of the 
theory.

Of course, though less natural, a theory can have different mass scales. 
Likewise, the internal symmetry breaking scale and the scale symmetry breaking
scale can be different. But, these scales must be built in someway.

\newsubsection{Effects of Scale Symmetry Breaking on Mass Scales}

The naturalness of mass scales in a theory has another interesting
implication. Consider spontaneous symmetry breaking involving two different
vacuum  expectations, which is a common situation in electroweak symmetry
breaking in supersymmetric models\cite{twohigg}\cite{revhig}\cite{susyrev}.
There is an undetermined parameter $\tan\beta = v_2/v_1$, where $v_1$ and $v_2$
are vacuum expectation values for two different scalars. It is very unnatural
to have two completely different dimensionful parameters\cite{dhiggs}. In the
presence of the dilaton, this unnaturalness can be accommodated 
even if it ever happens.

Let one of the scalar field have a scale-invariant mass term so that 
$\lambda_1\left(|\Phi_1|^2 - v^2\e^{2\kappa\phi}\right)^2$ for some reason
and the other have a scale-breaking mass term so that 
$\lambda_2\left(|\Phi_2|^2 - v^2\right)^2$.
Then symmetry breaking leads to $v_1 = v$ and 
$v_2 = v\e^{\kappa \langle\phi\rangle}$. This leads to 
$$\tan\beta = \e^{\kappa \langle\phi\rangle}.$$
Thus we only need one mass scale $v$. Note that, being exponential, $\tan\beta$
can be fairly large even for $\kappa \langle\phi\rangle \sim \CO(1)$.

Since we do not like to have an explicit scale-breaking term, in principle
the above could
be more realistically checked in the radiative case. In principle, the 
parameters to determine the two vacuum expectation values are 
coupling constants, the renormalization scale and the dilaton scale.
Upon scale symmetry breaking, the dilaton vacuum expectation value can 
eliminate the renormalization scale. Therefore, the two vacuum expectation
values should be related by the dilaton one, which always
enters exponentially.

\newsubsection{PCDC}

$\CL_{{\rm eff}}(\phi)$ in the new vacuum is a good candidate to address the
PCDC. 
After shifting the dilaton vacuum, the effective potential reads
\beq
\label{e44}
V_{\phi,{\rm eff}} = 
{\rm exp}\left\{2\left(1-{\pi^2\over \kappa^4\Lambda}\right)\right\}
{M^4\over 128\pi^2}\e^{4\kappa\phi}(4\kappa\phi -1).
\eeq
Let us first check the case of the classical scale symmetry $\delta M = 0$.
Under the variation, the leading scale symmetry breaking term is a constant
proportional to $M^4$. When this is translated into the dilatation current
conservation law, the leading anomalous term is constant. Thus it does not
fit to the idea of the PCDC, which implies that the dilatation current should
be conserved if the dilaton is massless, and any violation must be proportional
to the dilaton mass term.

Under the generalized scale symmetry, i.e. with eq.(\ref{e30}), we indeed have
the anomalous dilatation conservation law that meets the idea of the PCDC.
From the dilaton effective potential eq.(\ref{e44}), 
the generalized scale symmetry breaking leads to 
\beq
\label{eq21}
\pa_\mu S_{{\rm Q}}^\mu = -{m_\phi^2\over \kappa} \phi
\eeq
for small $\phi$, where the dilaton mass is given by
\beq
\label{eq22}
m^2_\phi = {\kappa^2 M^4\over 8\pi^2}\,
{\rm exp}\left\{2\left(1-{\pi^2\over \kappa^4\Lambda}\right)\right\}.
\eeq
Note that this dilaton mass is precisely the one arises in  $V_{\phi,{\rm
eff}}$ after shifting the vacuum as $\phi\to \phi + \langle\phi\rangle$, hence  
confirming that  our notion of spontaneous scale symmetry breaking is
consistent. The RHS of eq.(\ref{eq22}) is not really an anomaly, but takes an
analogous role here. 

Note that $m^2_\phi \ll \kappa^2 M^4$ for $\langle\phi\rangle \sim 1/\kappa$. 
In other words, the dilaton mass can be much smaller than the dilaton scale, 
giving us an expectation that the effect of scale symmetry breaking can be 
observed in much lower energy scale. This is much different from the Higgs
case,  where the Higgs mass is comparable to or even heavier than the
electroweak  symmetry breaking scale.

If the dilaton scale is the same as chiral symmetry breaking scale, say, 100
MeV, then the dilaton could be as light as a few KeV.  The dilaton could be a
candidate for dark matter. The dilaton couples to fermions and scalars quite
universally except the gauge fields, it could be abundant everywhere in the
universe. Thus it is worth while to further investigate physics of low energy
dilaton,  despite that there are not many scalar particles observed in the low
energy region.

Even if the low energy dilaton does not exist, it still does not mean 
the dilaton may cause a phenomenological disaster. The dilaton scale
could be very high, making the dilaton massive enough to decay into other 
particles rapidly at high energy.

\newsection{Conclusions}

There have been investigations to understand the relation between the scale 
symmetry in flat spacetime and the Weyl invariance in curved spacetime without
explicitly introducing the dilaton\cite{polch}\cite{rsw} and with the dilaton
as Brans-Dicke field\cite{bd}. In our case, we have shown that the scale
symmetry in flat space can be more elegantly described by the Diff symmetry of
curved spacetime. The dilaton is correctly identified only if SDiff is
incorporated. Being non-scalar, the dilaton is not the Brans-Dicke field. Since
scale invariant phenomenological Lagrangians can be derived naturally, we
believe this is the structure relevant to low energy dilatation physics. 

The dilatation current is derived and we have shown that there is an additional
term due to the dilaton, compared to the case without the dilaton. Conformal 
invariance is a natural consequence of scale invariance. This also proves that
two dimensional Liouville theory with the exponential potential term is indeed
a conformal field theory.

In this paper we focussed mainly on the implications of spontaneous scale
symmetry breaking in four dimensions. But we expect the same idea could be
accommodated in studying the quantum Liouville theory.

One subtlety we have not fully explained is if the 1PI effective potential we
derived is really unambiguously defined. The reason we believe so is that the
effective field theory  is assumed to be the low energy realization of a finite
or renormalizable theory. Although there are infinite orders of interactions
involved, a proper structure of counter terms might
exist to make the computation reasonable. 
Perhaps this could be checked more precisely by investigating any
cohomological structure of counter terms\cite{weinmorr}.

To break scale symmetry involving a dilaton vacuum expectation value is an
important task to accomplish in the context of string theory or supergravity.
We hope a supersymmetric generalization of the structure presented here  would
shed new light on that. The best way to supersymmetrize might be to use a four
dimensional analog to the super-Liouville theory. 

There are also much more works needed to realize the low energy dilaton in
nature. Just to name a few directions: The dilaton could be a source of dark
matter. There might be interesting dilaton-axion dynamics. Is there any 
relation between scale symmetry breaking and chiral symmetry breaking? Any
relevance in two-Higgs-doublet electroweak symmetry breaking? ...etc. In
particular, if the dilaton scale is the same as chiral symmetry breaking scale,
then one could think about generalizing chiral Lagrangians to incorporate the
dilaton. The simplest way is to dress $\Sigma = {\rm exp} (i\pi_a t^a)$ so that
$\Sigma_\chi \equiv \chi \Sigma$, which naturally reproduces the linear
sigma model Lagrangian. In the usual Goldstone-boson dynamics based on the
linear sigma model such a contribution of $\chi$ is often neglected in the
energy scale lower than the symmetry breaking scale. We now have a new
motivation  to look into this more carefully.

We hope the results obtained here would be helpful for the future progress on
dilatations. Further results will be presented soon.


{\renewcommand{\Large}{\large}

}

\end{document}